\documentstyle[prd,preprint,tighten,aps,eqsecnum,amssymb]{revtex}
\begin{document}
\draft
\preprint{
\begin{tabular}{r}
DFTT 54/94
\\
SNUTH-94-109
\\
hep-ph/9411219
\end{tabular}
}
\title{Atmospheric Neutrino Problem in Maximally-Mixed
\\
Three Generations of Neutrinos}
\author{
C. Giunti$^{\mathrm{a}}$\thanks{E-mail address: GIUNTI@TO.INFN.IT.},
C. W. Kim$^{\mathrm{a,b}}$\thanks{On leave of absence from
The Johns Hopkins University,
Baltimore, Maryland 21218, USA.
E-mail address: CWKIM@JHUVMS.HCF.JHU.EDU.}
and
J. D. Kim$^{\mathrm{b}}$\thanks{E-mail address: JDKIM@PHYB.SNU.AC.KR.}
}
\address{
\begin{tabular}{c}
$^{\mathrm{a}}$INFN, Sezione di Torino
and Dipartimento di Fisica Teorica, Universit\`a di Torino,
\\
Via P. Giuria 1, 10125 Torino, Italy
\\
$^{\mathrm{b}}$Department of Physics
and Center for Theoretical Physics,
\\
Seoul National University, Seoul, Korea
\end{tabular}
}
\date{March 1995}
\maketitle
\begin{abstract}
Motivated by the indication that both the atmospheric and the
solar neutrino puzzles may simultaneously be solved by
(vacuum as well as matter-induced
resonant) oscillations of two generations of neutrinos with large mixing,
we have analyzed the data on the atmospheric and solar neutrinos
assuming that all {\it three} neutrinos are maximally mixed.
It is shown that the
values of
$ \Delta m^2 $
obtained from the
two-generation analyses
are still valid even in the three-generation scheme,
i.e.
the two puzzles can be solved simultaneously if
$ \Delta m_{31}^2 \simeq 10^{-2} \, \mathrm{eV}^2 $
for the atmospheric neutrinos
and
$ \Delta m_{21}^2 \simeq 10^{-10} \, \mathrm{eV}^2 $
for solar neutrinos
in the maximally mixed three-generation scheme.
\end{abstract}



\narrowtext

\section{Introduction}
\label{S:INTRO}

One of the several outstanding issues that we have faced so long
in particle
physics is that of neutrino masses and mixing
(see, for example, Refs.\cite{B:BILENKY,B:CWKIM}).
Studies of solar neutrinos have been very
promising for some time and already we have preliminary indications
for positive answers
\cite{B:HOMESTAKE,B:KAMIOKANDE,B:GALLEX,B:SAGE}.
In this case,
however,
one has to rely on the solar model
\cite{B:BAHCALL},
which is as good as it can be but it may not be,
as yet,
perfect.
For this reason,
many suggestions have been
proposed on solar model-independent
ways of analyzing solar neutrino
data when more data will become available in the future
\cite{B:MIT,B:BG}.

Another prominent source of
neutrinos that has been
investigated in the past is
atmospheric neutrinos
\cite{B:KAM88,B:KAM92,B:KAM94,B:IMB,B:FREJUS,B:NUSEX,B:SOUDAN}.
In this case,
although the
absolute fluxes of
$\nu_{\mu}$ ($\bar\nu_{\mu}$)
and
$\nu_{e}$ ($\bar\nu_{e}$)
are uncertain up to 30\%
\cite{B:GS},
if one takes
the ratio of neutrino-induced muon and electron events both for data and MC
simulations
(in fact the ratio of the ratios),
the results are almost model independent with errors of about 5\%
\cite{B:GS}.
In this sense,
the atmospheric neutrinos may be better suited for
extracting information on the neutrino
parameters such as mass and mixing angle.

Recently the Kamiokande collaboration
\cite{B:KAM94}
has analyzed more
than 200 events with
energies in the multi-GeV range and
updated the previous sub-GeV data.
Based on these data they have concluded that both data can
be explained by either
$\nu_{e}$-$\nu_{\mu}$
or
$\nu_{\mu}$-$\nu_{\tau}$
oscillations.
However,
they have used explicitly the two generation formulations.
The
results of their analyses under the assumption that the anomalies are
due to neutrino oscillations  are
summarized as being consistent with
$ \sin^2(2\vartheta) \simeq 1 $
and
$ \Delta m^2 \simeq 10^{-2} $,
where $\vartheta$ and
$\Delta m^2$ are,
respectively,
the $\nu_{e}$-$\nu_{\mu}$
mixing angle
and
$ m_2^2 - m_1^2 $,
and
the $\nu_{\mu}$-$\nu_{\tau}$
mixing angle
and
$ m_3^2 - m_2^2 $
for the
$ \nu_{e} \leftrightarrows \nu_{\mu} $
and
$ \nu_{\mu} \leftrightarrows \nu_{\tau} $
interpretations.
(Whenever we mention
$\nu_{e}$,
$\nu_{\mu}$
and
$\nu_{\tau}$
we also infer the cases of
$\bar\nu_{e}$,
$\bar\nu_{\mu}$
and
$\bar\nu_{\tau}$.)

We also recall that in the
case of solar neutrinos
there are basically two popular ways to
explain
the so-called anomaly
(under the assumption that the SSM is correct) with
neutrino oscillations,
one with vacuum oscillations
and the other with the MSW effect
\cite{B:MSW}.
Both cases contain
solutions with almost maximal mixing of
the neutrinos involved
\cite{B:SNPVO,B:SNPMSW}.
It is to be stressed here
that for reaching the respective conclusions
in all these analyses
two generations of neutrinos have been used.

Now we may ask an obvious and logical question:
Do these
results remain valid even when the third generation neutrino is
included?
After all,
we do know that there exist three
neutrinos associated with $e$, $\mu$ and $\tau$.
However,
a three-generation analysis of any phenomenon is known to be very
complicated and tedious,
and often it is hard to see the physics involved
\cite{B:3G}.

In this work, we analyze the atmospheric
and solar neutrinos using the three generations of neutrinos
that are maximally
mixed to begin with.
Although this is,
admittedly, an extreme case, it is a very
good approximation in the event of large neutrino mixings,
and,
moreover,
the algebra and
analysis of the data become
very simple and the physics becomes
transparent.
Furthermore, this does not imply a loss of generality
if mixings turn out to be reasonably large
because
the qualitative results still remain valid.

The three-generation neutrino oscillation with maximal mixing has
also been
discussed in the past
\cite{B:ALPW93}
without taking into account the energy
and distance dependence of
the oscillating terms.
In this case the survival probability for the solar neutrinos,
$P(\nu_e \to \nu_e)$,
is 1/3
at all energies.
In Ref.\cite{B:ALPW93}
also the atmospheric neutrino anomaly has been
addressed by assuming that all
$\Delta m_{ij}^2$
are in the range
$ (0.5 - 1.2) \times 10^{-2} \mathrm{eV}^2 $.

Our main purpose in this paper is to see if and how the two generation
results are  modified when the third neutrino generation is introduced
by taking into account
the energy and distance dependence of the oscillating terms.

\section{Maximal Mixing among Three Neutrinos}
\label{S:MAXMIX}

We consider three neutrinos to be maximally mixed
if their survival probabilities are all equal
and for distances much larger than the oscillation lengths
all the three survival probabilities average to $1/3$.
This can be accomplished with the following mixing matrix
\cite{B:NUSSINOV76,B:WOLFENSTEIN78}:
\begin{equation}
U
=
{1\over\sqrt{3}}
\left(
\begin{array}{ccc}
1 & 1 & 1 \\
1 & \mathrm{e}^{ 4 \pi i / 3 } & \mathrm{e}^{ 2 \pi i / 3 } \\
1 & \mathrm{e}^{ 2 \pi i / 3 } & \mathrm{e}^{ 4 \pi i / 3 }
\end{array}
\right)
\;.
\label{E03}
\end{equation}
The presence in $U$ of the cube roots of unity
$1$,
$\mathrm{e}^{ 2 \pi i / 3 }$,
$\mathrm{e}^{ 4 \pi i / 3 }$
is necessary for $U$ to be unitary.
It follows that in the maximal mixing case
CP must be violated.
Using the mixing matrix (\ref{E03})
we obtain the survival probabilities
\begin{eqnarray}
P(\nu_e \to \nu_e)
& = &
P(\nu_\mu \to \nu_\mu)
\; = \;
P(\nu_\tau \to \nu_\tau)
\nonumber
\\
&=&
1
-
{4\over9}
\left(
\sin^2 k_{21}
+
\sin^2 k_{31}
+
\sin^2 k_{32}
\right)
\;,
\label{E040}
\end{eqnarray}
with
\begin{equation}
k_{ij} \equiv \frac{1.27L\Delta m_{ij}^2}{E}
\;.
\end{equation}
Here
$L$, $E$ and
$ \Delta m_{ij}^2 \equiv m_i^2-m_j^2 $
are in units of meter, MeV and $\mathrm{eV}^2$,
respectively.
On the other hand,
the probabilities of transition
between different neutrino flavors are not identical.
Instead,
we have
\begin{eqnarray}
P(\nu_{\mu}\to\nu_{e})
& = &
P(\nu_{e}\to\nu_{\tau})
\; = \;
P(\nu_{\tau}\to\nu_{\mu})
\nonumber
\\
&=&
1
-
{4\over9}
\left[
\sin^2
\left(
k_{21}
+
{\displaystyle
\pi
\over\displaystyle
3
}
\right)
+
\sin^2
\left(
k_{31}
-
{\displaystyle
\pi
\over\displaystyle
3
}
\right)
+
\sin^2
\left(
k_{32}
+
{\displaystyle
\pi
\over\displaystyle
3
}
\right)
\right]
\label{E018}
\end{eqnarray}
and
\begin{eqnarray}
P(\nu_{e}\to\nu_{\mu})
& = &
P(\nu_{\mu}\to\nu_{\tau})
\; = \;
P(\nu_{\tau}\to\nu_{e})
\nonumber
\\
&=&
1
-
{4\over9}
\left[
\sin^2
\left(
k_{21}
-
{\displaystyle
\pi
\over\displaystyle
3
}
\right)
+
\sin^2
\left(
k_{31}
+
{\displaystyle
\pi
\over\displaystyle
3
}
\right)
+
\sin^2
\left(
k_{32}
-
{\displaystyle
\pi
\over\displaystyle
3
}
\right)
\right]
\label{E019}
\end{eqnarray}

It is interesting to notice that
the mixing matrix for maximal mixing is not
unique.
In fact,
all the matrices which are obtained from $U$
by multiplying any number of rows or columns
by a phase give the same oscillation probabilities
(\ref{E040}), (\ref{E018}) and (\ref{E019}).
Furthermore,
$U^{\dagger}$
gives the same survival probabilities
(\ref{E040}),
but all the phases
$ \pi / 3 $
in the transition probabilities
(\ref{E018}) and (\ref{E019})
change sign.
Since the definition of maximal mixing
requires the survival probabilities
(\ref{E040}),
it follows that both
$U$ and $U^{\dagger}$
give maximal mixing,
but they produce two different
phenomenologies for the transition probabilities.

\section{Atmospheric Neutrinos}
\label{S:ATM}

As in many previous works,
our study of atmospheric neutrinos
involves the ratio
$ \left( \nu_{\mu} + \bar\nu_{\mu} \right)
/ \left( \nu_{e} + \bar\nu_{e} \right) $.
In actual experimental situations,
this ratio manifests as an observed ratio
of muon-like and electron-like events.
Although the issue of whether or not
the observed ratio
is indeed small compared with
the Monte Carlo (MC)
simulated ratio is still not completely settled
\cite{B:KAM88,B:KAM92,B:KAM94,B:IMB,B:FREJUS,B:NUSEX,B:SOUDAN},
the Kamiokande group recently presented
two sets of impressive data,
one involving the electron and muon events
in the sub-GeV energy range
and the other in the multi-GeV energy range.
The Kamiokande collaboration found
\cite{B:KAM94}
\begin{equation}
\mathrm{R}^{\mathrm{exp}}
\equiv
{\displaystyle
\mathrm{R} \left( \mu / e \right)_{\mathrm{exp}}
\over\displaystyle
\mathrm{R} \left( \mu / e \right)_{\mathrm{MC}}
}
=
\left\{
\begin{array}{rcl} \displaystyle
0.60^{+0.06}_{-0.05} \pm 0.05
\null & \null \hskip0.5cm \null & \null \displaystyle
\mbox{for the sub-GeV energy range}
\; ,
\\ \displaystyle
\null & \null \hskip0.5cm \null & \null \displaystyle
\\ \displaystyle
0.57^{+0.08}_{-0.07} \pm 0.07
\null & \null \hskip0.5cm \null & \null \displaystyle
\mbox{for the multi-GeV energy range}
\; ,
\end{array}
\right.
\end{equation}
where
$ \mathrm{R} \left( \mu / e \right)_{\mathrm{exp}} $
and
$ \mathrm{R} \left( \mu / e \right)_{\mathrm{MC}} $
are the ratios of
$\mu$-like and $e$-like
events observed and calculated,
respectively.
In addition,
they studied the dependence of the above ratios
on the zenith-angle $\theta$
in the range
$ -1 \le \cos\theta \le 1 $,
divided in 5 bins.
The zenith-angle dependence of $\mathrm{R}^{\mathrm{exp}}$
for the sub-GeV data
is rather flat
(see Fig.6 of Ref.\cite{B:KAM94}),
whereas
that of the multi-GeV data
appears to deviate from being flat
(see Fig.4 of Ref.\cite{B:KAM94}).
Our analysis will be based only on the Kamiokande data.

Let us consider
the zenith-angle dependence of the
Kamiokande data.
For each bin $i$
($i=1,\ldots,5$
for
$ \left\langle \cos\theta \right\rangle_{i=1,\ldots,5} =
-0.8 \, , \, -0.4 \, , \, 0.0 \, , \, 0.4 \, , \, 0.8 $,
which correspond to the averaged distances
$ \left\langle L \right\rangle_{i=1,\ldots,5} =
10,230 \, , \, 5,157 \, , \, 852 \, , \, 54 \, , \, 26 \, \mathrm{Km} $),
the number
$ N_{e i} $
of $e$-like events,
the number
$ N_{\mu i} $
of $\mu$-like events
and
the ratio of ratios
$\mathrm{R}_{i}$
are given by
\begin{eqnarray}
&&
N_{e i}
=
{ \rho_{\mathrm{T}} \over \sqrt{ \rho_{\mathrm{R}} } }
\,
N_{e i}^{\mathrm{MC}}
P_{i}(\nu_{e}\to\nu_{e})
+
\rho_{\mathrm{T}} \, \sqrt{ \rho_{\mathrm{R}} }
\,
N_{\mu i}^{\mathrm{MC}}
\,
P_{i}(\nu_{\mu}\to\nu_{e})
\; ,
\label{E011}
\\
&&
N_{\mu i}
=
{ \rho_{\mathrm{T}} \over \sqrt{ \rho_{\mathrm{R}} } }
\,
N_{e i}^{\mathrm{MC}}
\,
P_{i}(\nu_{e}\to\nu_{\mu})
+
\rho_{\mathrm{T}} \, \sqrt{ \rho_{\mathrm{R}} }
\,
N_{\mu i}^{\mathrm{MC}}
P_{i}(\nu_{\mu}\to\nu_{\mu})
\;,
\label{E012}
\\
&&
\mathrm{R}_{i}
\equiv
{\displaystyle
N_{\mu i} / N_{\mu i}^{\mathrm{MC}}
\over\displaystyle
N_{e i} / N_{e i}^{\mathrm{MC}}
}
=
\rho_{\mathrm{R}}
\,
{\displaystyle
P_{i}(\nu_{\mu}\to\nu_{\mu})
+
\rho_{\mathrm{R}}^{-1}
\left( N_{\mu i}^{\mathrm{MC}} / N_{e i}^{\mathrm{MC}} \right)^{-1}
P_{i}(\nu_{e}\to\nu_{\mu})
\over\displaystyle
P_{i}(\nu_{e}\to\nu_{e})
+
\rho_{\mathrm{R}}
\left( N_{\mu i}^{\mathrm{MC}} / N_{e i}^{\mathrm{MC}} \right)
P_{i}(\nu_{\mu}\to\nu_{e})
}
\; .
\label{E013}
\end{eqnarray}
Here
$ N_{e i}^{\mathrm{MC}} $
and
$ N_{\mu i}^{\mathrm{MC}} $
are,
respectively,
the number
of $e$-like and $\mu$-like events
in the bin $i$
predicted by MC simulations,
without neutrino oscillations,
after passing through the same analysis chain as the data.
The factors
$ \rho_{\mathrm{T}} $
and
$ \rho_{\mathrm{R}} $
are the normalization coefficients
that take into account the
systematic error of the MC calculation of the
total number of
$e$-like and $\mu$-like events (30\%)
and the
systematic error of the MC calculation of the
$\mu/e$ ratio
(9\% for the sub-GeV data
and
12\% for the multi-GeV data),
respectively.
We inserted these two factors
in Eqs.(\ref{E011}) and (\ref{E012})
in such a way that
the ratio of ratios
$ \mathrm{R}_{i} $
does not depend on
$ \rho_{\mathrm{T}} $,
i.e.,
only the coefficient
$ \rho_{\mathrm{R}} $
appears in the expression (\ref{E013})
for the ratio of ratios
as an overall multiplicative factor
and
the ratio
$ \left( N_{\mu i}^{\mathrm{MC}} / N_{e i}^{\mathrm{MC}} \right) $
is always multiplied by
$ \rho_{\mathrm{R}} $.
As a consequence,
the error of
$ \mathrm{R}_{i} $
does not depend on the
systematic error of the MC calculation of the total number of
$e$-like and $\mu$-like events,
but only
on the systematic error of the MC calculation of the
$\mu/e$ ratio,
which is much smaller.
In Eqs.(\ref{E011})--(\ref{E013})
we neglected the fact that not all
$e$-like and $\mu$-like events
are produced by
$ \nu_{e} $ ($\bar\nu_{e}$)
and
$\nu_{\mu} $ ($\bar\nu_{\mu}$),
respectively.
However,
the purity of the
$ \nu_{e} $ and $\nu_{\mu} $
contributions to the selected
$e$-like and $\mu$-like events
is estimated by the Kamiokande Collaboration
to be higher than 90\%.
We wish to emphasize here
that our three generation formalism
takes into account simultaneously
the two neutrino oscillation channels
$ \nu_{\mu} \leftrightarrows \nu_{e} $
and
$ \nu_{\mu} \to \nu_{\tau} $.
This is not the case for two generation analyses,
where the two oscillation channels
are considered separately,
leading to two possible
interpretations and results,
as done in Ref.\cite{B:KAM94}.

Since we have
$ L \lesssim 13 \times 10^3 \, \mathrm{Km} $,
$ E \gtrsim 0.2 \, \mathrm{GeV} $ and
$ \Delta m_{21}^2 \simeq 10^{-10} \, \mathrm{eV}^2 $
(as will be seen later
from the solution of the solar neutrino problem),
$k_{21}$ is well
approximated by zero and,
because of the assumed hierarchy $m_3>m_2>m_1$,
we expect
$ k_{32} \simeq k_{31} $,
so that the relevant transitions
probabilities are
\begin{equation}
P(\nu_e\to\nu_\mu)
=
P(\nu_\mu\to\nu_e)
=
P(\nu_\mu \to \nu_\tau)
=
P(\nu_e\to\nu_\tau)
=
{\displaystyle 4\over\displaystyle 9}
\,
\sin^2k_{32}
\;.
\label{E:14}
\end{equation}

We obtained the values of
$ N_{e i}^{\mathrm{MC}} $
and
$ N_{\mu i}^{\mathrm{MC}} $
from the MC data
for the $e$-like and $\mu$-like events
presented
in Fig.2 of Ref.\cite{B:KAM88}
for the sub-GeV data
and the MC data presented
in Fig.3 of Ref.\cite{B:KAM94}
for the multi-GeV data.
The quantities
$ P_{i}(\nu_{e}\to\nu_{\mu}) $
and
$ P_{i}(\nu_{\mu}\to\nu_{e}) $
are the
$ \nu_{e}\to\nu_{\mu} $
and
$ \nu_{\mu}\to\nu_{e} $
transition probabilities
averaged over the neutrino energy spectrum
and the zenith-angle width of bin $i$.
These probabilities depend on the value of
$ \Delta m^2_{32} $
and
they are different because
initial $ \nu_{e} $ and $ \nu_{\mu} $
have different energy spectra.
These energy spectra are given
in Fig.1 of Ref.\cite{B:KAM92}
for the sub-GeV data
and the MC data presented
in Fig.2 of Ref.\cite{B:KAM94}
for the multi-GeV data.

Before we present the results of our full analysis,
we discuss the following simple but interesting possibility.
As suggested by the zenith-angle distribution
of the sub-GeV data
and to some extent
(within error bars)
by the large-angle data points
of the multi-GeV data
with $ i=1,\ldots,4 $,
corresponding to
$ \left\langle \cos\theta \right\rangle $
from -0.8 to 0.4,
$\mathrm{R}^{\mathrm{exp}}$
is roughly consistent with being flat.
This may be interpreted
as due to washing out of
the distance dependence
in the argument of
$ \sin^2 \left( 1.27 L \Delta m^2_{32} / E \right) $
(the averaged distance $ \left\langle L \right\rangle $
ranges from
10,230 Km to 26 Km for the sub-GeV data
and from
10,230 Km to 54 Km for the multi-GeV data).
That is,
the oscillations are in the rapid oscillation region
with
$ \sin^2 \left( 1.27 L \Delta m^2_{32} / E \right) \simeq 1/2 $
and
$ P(\nu_{\alpha}\to\nu_{\beta}) \simeq 2/9 $
for $ \alpha \not= \beta $.
Under this assumption,
Eq.(\ref{E013})
reduces to
\begin{equation}
\mathrm{R}_{i}
=
\rho_{\mathrm{R}}
\,
{\displaystyle
1
-
{1\over9}
+
{2\over9}
\,
\rho_{\mathrm{R}}^{-1}
\left( N_{\mu i}^{\mathrm{MC}} / N_{e i}^{\mathrm{MC}} \right)^{-1}
\over\displaystyle
1
-
{1\over9}
+
{2\over9}
\,
\rho_{\mathrm{R}}
\left( N_{\mu i}^{\mathrm{MC}} / N_{e i}^{\mathrm{MC}} \right)
}
\; .
\label{E002}
\end{equation}
Since we have
$ \left. N_{\mu i}^{\mathrm{MC}} / N_{e i}^{\mathrm{MC}} \right|_{i=1,\ldots,5}
\simeq 1.57 $
for the sub-GeV data
and
$ \left. N_{\mu i}^{\mathrm{MC}} / N_{e i}^{\mathrm{MC}} \right|_{i=1,\ldots,4}
\simeq 2.98 \, , \, 2.29 \, , \, 2.08 \, , \, 2.39 $
for the multi-GeV data,
and taking into account the systematic error of
$ \rho_{\mathrm{R}} $,
we find
$ \mathrm{R}_{i=1,\ldots,5} = 0.83 \pm 0.04 $
for the sub-GeV data
and
$ \mathrm{R}_{i=1,\ldots,4} =
0.62 \pm 0.04
\, , \,
0.70 \pm 0.05
\, , \,
0.74 \pm 0.05
\, , \,
0.69 \pm 0.04 $
for the multi-GeV data.
These values are to be
compared with
$ \left\langle \mathrm{R}_{i}^{\mathrm{exp}} \right\rangle =
0.60^{+0.08}_{-0.07} $
for the average of the sub-GeV data
and
$ \mathrm{R}_{i=1,\ldots,4}^{\mathrm{exp}} =
0.29^{+0.12}_{-0.07}
\, , \,
0.46^{+0.17}_{-0.13}
\, , \,
0.51^{+0.15}_{-0.11}
\, , \,
0.63^{+0.22}_{-0.16}
$
for the multi-GeV data.
It appears that
the idea of rapid oscillations is not too
inconsistent with the observations.

\begin{figure}
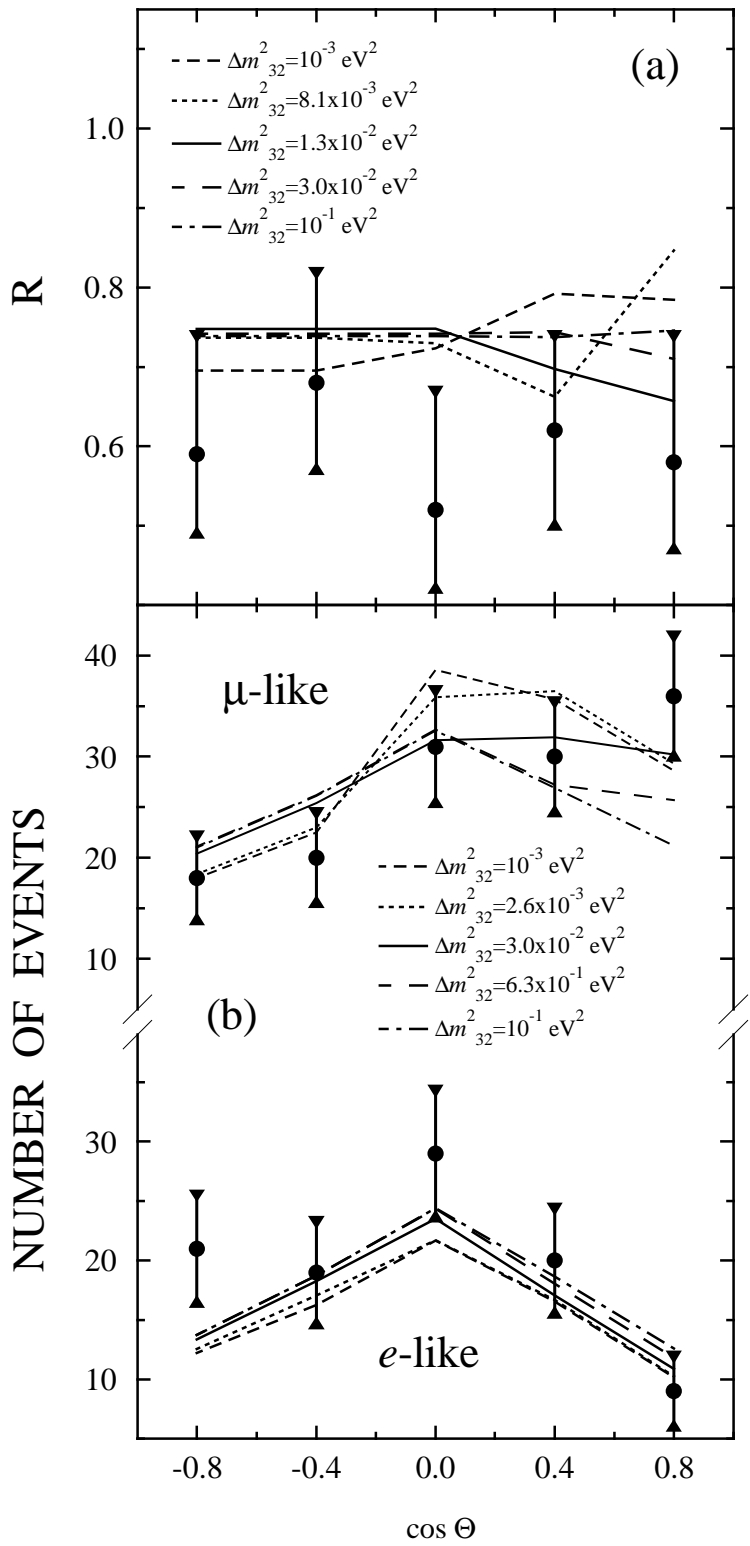

\caption{Separate fits of the sub-GeV (a)
and multi-GeV (b) data
for several different values of $ \Delta m^2_{32} $.}
\label{F:FIG1}
\end{figure}

\begin{figure}
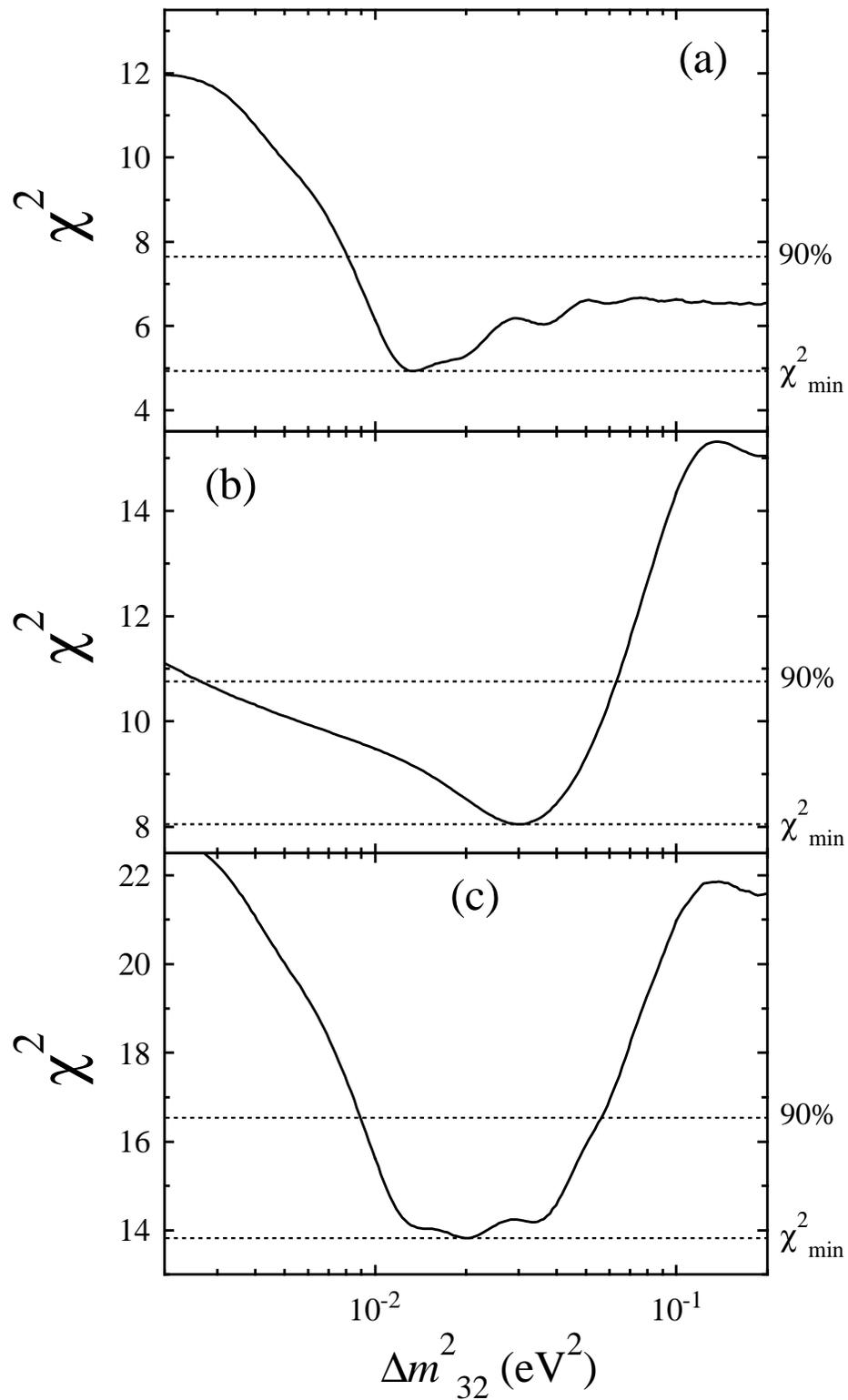

\caption{Values of $\chi^2$
for the separate fits of the sub-GeV (a)
and multi-GeV (b) data
and the combined fit (c).}
\label{F:FIG2}
\end{figure}

Let us now discuss
the results of a full analysis
of the Kamiokande data
taking into account the oscillation effect.
For different values of $ \Delta m^2_{32} $
we have fitted separately
the sub-GeV data of the ratio of ratios
presented in Fig.6 of Ref.\cite{B:KAM94}
and
the multi-GeV data
of the number of $e$-like and $\mu$-like events
presented in Fig.3 of Ref.\cite{B:KAM94}.
For the fit of the sub-GeV data
we used the theoretical formula
for the ratio of ratios
given in Eq.(\ref{E013})
and
for the fit of the multi-GeV data
we used the theoretical formulas
for the number of $e$-like and $\mu$-like events
given in Eqs.(\ref{E011}) and (\ref{E012}).
For each value of $ \Delta m^2_{32} $
we have calculated the minimum value of the $\chi^2$
with respect to the normalization coefficients
$ \rho_{\mathrm{T}} $
and
$ \rho_{\mathrm{R}} $,
whose difference from unity
contributes to the $\chi^2$
with errors
$ \sigma(\rho_{\mathrm{R}}) = 9\% $
for the sub-GeV data
and
$ \sigma(\rho_{\mathrm{T}}) = 30\% $,
$ \sigma(\rho_{\mathrm{R}}) = 12\% $
for the multi-GeV data.
Some results are depicted in Fig.\ref{F:FIG1}
together with the experimental data.
As can be seen in Fig.\ref{F:FIG1},
the sub-GeV data and multi-GeV data
can be best fitted (solid lines)
with
$ \Delta m^2_{32} = 1.3 \times 10^{-2} \, \mathrm{eV}^2 $
and
$ \Delta m^2_{32} = 3.0 \times 10^{-2} \, \mathrm{eV}^2 $,
respectively.
In Fig.\ref{F:FIG2}
we have plotted the $\chi^2$ of the fits
as functions of
$ \Delta m^2_{32} $.
As mentioned already,
the best fits of
the sub-GeV
(Fig.\ref{F:FIG2}a)
and multi-GeV
(Fig.\ref{F:FIG2}b)
data
are obtained for
$ \Delta m^2_{32} = 1.3 \times 10^{-2} \, \mathrm{eV}^2 $
($ \chi^2 = 4.9 $ with 3 degrees of freedom)
and
$ \Delta m^2_{32} = 3.0 \times 10^{-2} \, \mathrm{eV}^2 $
($ \chi^2 = 8.1 $ with 7 degrees of freedom),
respectively.
The separate fits of the
the sub-GeV
and multi-GeV data
are acceptable at
18\% CL and 32\% CL,
respectively.
The 90\% CL allowed range of
$ \Delta m^2_{32} $
is determined by
$ \chi^2 < \chi^2_{\mathrm{min}} + 2.7 $,
that gives
$ \Delta m^2_{32} > 8.1 \times 10^{-3} \, \mathrm{eV}^2 $
from the sub-GeV data
and
$ 2.6 \times 10^{-3} \, \mathrm{eV}^2
< \Delta m^2_{32}
< 6.3 \times 10^{-2} \, \mathrm{eV}^2 $
from multi-GeV data.
We see that these allowed ranges are
perfectly compatible.
Fig.\ref{F:FIG2}c
gives the $\chi^2$ of the combined fit
of the sub-GeV
and multi-GeV data.
Here there is a minimum for
$ \Delta m^2_{32} = 2.0 \times 10^{-2} \, \mathrm{eV}^2 $,
very close to the two independent best fits
of the sub-GeV
and multi-GeV data.
Since
$ \chi^2_{\mathrm{min}} = 13.8 $
and we have 11 degrees of freedom,
the fit is acceptable at 24\% CL.
As can be seen from
Fig.\ref{F:FIG2}c,
the simultaneous fit
of both the sub-GeV
and multi-GeV data
allows to constrain the value of
$ \Delta m^2_{32} $
to lie in the interval
$ 8.9 \times 10^{-3} \, \mathrm{eV}^2
  < \Delta m^2_{32} <
  5.6 \times 10^{-2} \, \mathrm{eV}^2 $
at 90\% CL.
This result agrees well
with those given in Ref.\cite{B:KAM94},
obtained from two generation analyses.
We wish to add here
that the values of
$ \Delta m^2_{32} $
larger than
$ 9.0 \times 10^{-3} \, \mathrm{eV}^2 $
and larger than
$ 2.0 \times 10^{-2} \, \mathrm{eV}^2 $
are excluded at 90\% CL
by the
Kurchatov
\cite{B:KURCHATOV90}
and
Gosgen
\cite{B:GOSGEN86}
reactor neutrino oscillation experiments,
respectively.
If combined,
we are left with a very
restricted allowed range for the value of
$ \Delta m^2_{32} $,
which is about
$ 10^{-2} \, \mathrm{eV}^2 $.

\section{Solar Neutrinos}
\label{S:SOL}

We now proceed to check
if the three-generation maximally mixed
scheme discussed so far
is also capable of solving
the solar neutrino problem.
Since we assume
$ \Delta m_{31}^2 \simeq \Delta m_{32}^2 \gg 10^{-10} \, \mathrm{eV}^2$
as obtained in
the atmospheric neutrino problem,
we have,
from Eqs.(\ref{E018}) and (\ref{E019}),
\begin{equation}
{\displaystyle
P(\nu_e \to \nu_\mu)
+
P(\nu_e \to \nu_\tau)
\over\displaystyle
2
}
=
{2\over9}
+
{2\over9}
\,
\sin^2 k_{21}
\;,
\label{E:ELTOMU}
\end{equation}
and,
from Eq.(\ref{E040}),
\begin{equation}
P(\nu_e \to \nu_e)
=
{5\over9}
-
{4\over9}
\sin^2 k_{21}
\;.
\label{E:ELTOEL}
\end{equation}
Note that the maximal value of the survival probability of $\nu_e$
is $\left(5/9\right)=0.56$.
This implies that, since the Standard Solar Model predicts the event
rate for the Gallium detectors
to be $132^{+9}_{-6}$ SNU, the observed event rate must be
\begin{equation}
\Sigma_{\mathrm{exp}} \lesssim 80\ .
\label{E:gap}
\end{equation}
If the $\Sigma_{\mathrm{exp}}$
turns out to be greater than the above, this maximal mixing model
(as an explanation with vacuum oscillation) is ruled out (or
$\Delta m_{32}^2$
and
$\Delta m_{31}^2$
must take values such that
$\sin^2k_{32}$
and
$\sin^2k_{31}$
cannot be approximated by 1/2).
The latest average of the
GALLEX \cite{B:GALLEX}
and
SAGE \cite{B:SAGE}
event rates is
$ \Sigma_{\mathrm{exp}} = 74 \pm 9 $
which is still consistent with Eq. (\ref{E:gap}).

\begin{figure}
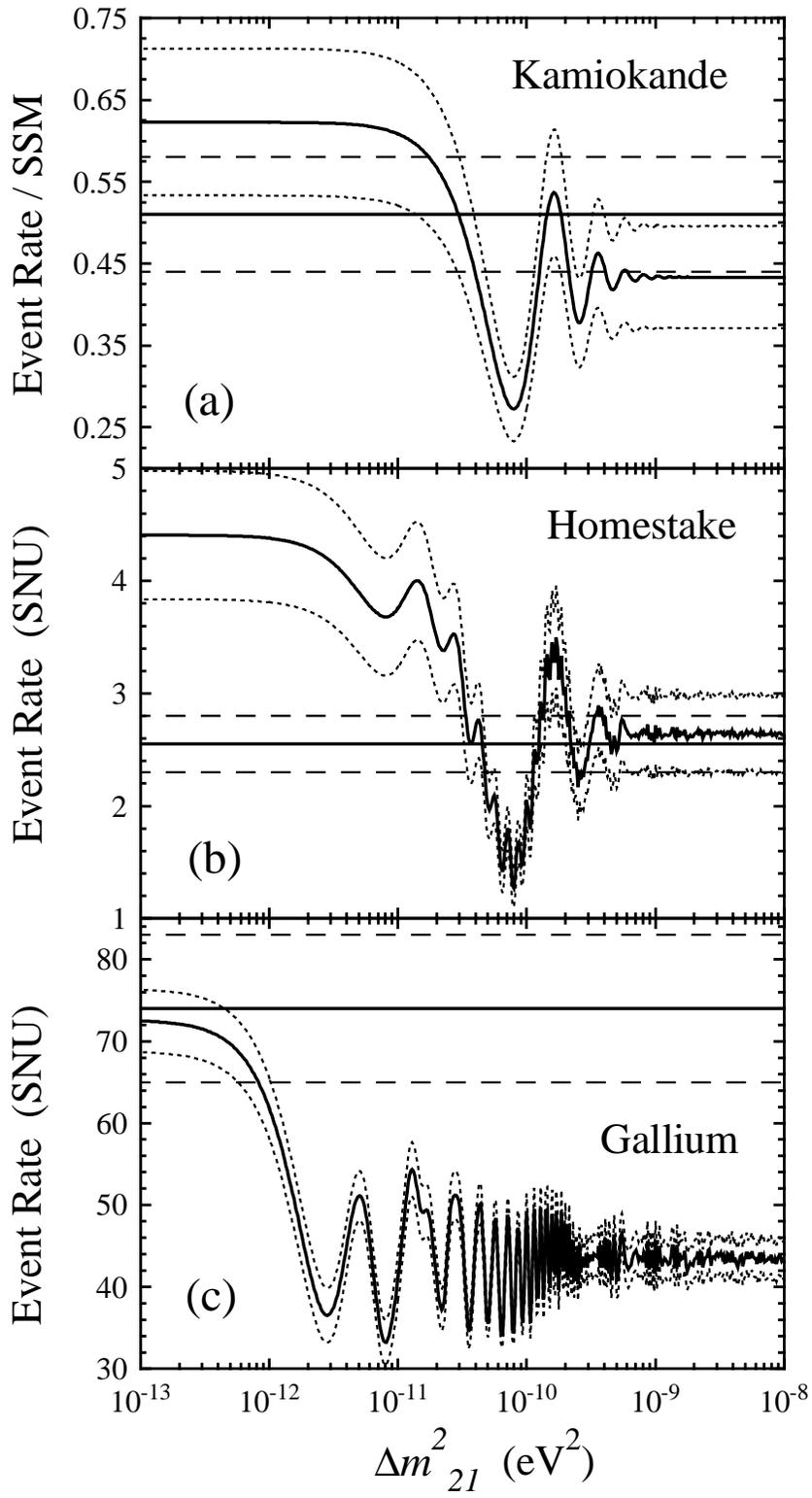

\caption{Solar neutrino
event rates as functions
of $\Delta m^2_{21}$.
The horizontal solid lines are the central values of the
experiments with the dashed lines representing one standard deviation.
The oscillating solid lines represent the theoretical values,
with the dotted lines indicating one
standard deviation resulting from the errors in the SSM.}
\label{F:FIG3}
\end{figure}

The results of the calculations
for the Kamiokande,
Homestake and Gallium experiments
are shown in Fig.\ref{F:FIG3}.
These calculations
are based on the standard procedure,
with the neutrino fluxes
predicted by the Bahcall and Pinsonneault SSM
\cite{BPSSM92},
the detector cross sections
given in Refs.\cite{B:BAHCALL,B:AUFDERHEIDE94},
and the probabilities given in
Eqs.(\ref{E:ELTOMU}) and (\ref{E:ELTOEL}).

In the figures, the horizontal solid lines are the central values of the
experiments with the dashed lines representing one standard deviation.
The oscillating solid lines represent the theoretical values,
with the dotted lines indicating one
standard deviation resulting from the errors in the SSM.
We can see that in the neighborhood of
$ \Delta m_{21}^2 \simeq 10^{-10} \, \mathrm{eV}^2 $
theory and experiments agree reasonably well,
with the exception
of the Gallium experiments,
in which case an agreement is obtained roughly
within two standard deviations.
The allowed values close to
$ \Delta m^2_{21} \simeq 10^{-10} \, \mathrm{eV}^2 $
are in agreement with those
obtained by two-generation analyses
\cite{B:ALPW93,B:SNPVO}.

\section{Conclusions}
\label{S:CONCL}

Assuming maximally-mixed
three-generations of neutrinos,
we have analyzed the recent data
on atmospheric neutrinos
obtained by the Kamiokande Collaboration,
which presented
a corresponding analysis in the
two-neutrino scheme.

The study is motivated by the indications
of possible large mixing angle solutions
for both the atmospheric and solar neutrino problems.
In particular,
in the recent Kamiokande analysis of the
atmospheric neutrino data,
the best fit values of
$ \sin^2 ( 2 \vartheta ) $
are shown to be almost unity.
Both the sub-GeV and multi-GeV data
can be explained
with
$ \Delta m^2_{32} \simeq 2 \times 10^{-2} \, \mathrm{eV}^2 $.
In our three-generation case
the range of values of
$ \Delta m^2_{32} $
agrees with that of the two-generation analyses,
but the values of
$ \sin^2 ( 2 \vartheta_{\alpha\beta} ) $
are 0.89
($ \alpha \, , \, \beta = e \, , \, \mu \, , \, \tau $)
because the maximal mixing angles
are $35^{\circ}$
instead of $45^{\circ}$,
which is the case in the two-generation schemes.
We conclude therefore that,
besides this difference,
the addition of the third generation of neutrinos
does not change the range of
$ \Delta m^2_{32} $
obtained in the two-generation analysis.

Our solutions are based on vacuum oscillations.
For the range of
$ \Delta m^2_{32} $
that we obtained,
the medium effects of the Earth are negligible.
Downward-going neutrinos
do not travel enough in matter
and
for
upward-going neutrinos
the large neutrino mixing
and the value of $ \Delta m^2_{32} $
make the MSW effect inoperative.

We have also checked the solar neutrino problem
in the same scheme.
One definite prediction
of the maximally mixed three-generation model
is that the observed event rate for the Gallium experiments
cannot be larger than 80 SNU.
With exception of the Gallium case,
the scheme provides
a good fit
of the data,
assuming vacuum oscillations
with values of
$ \Delta m^2_{21} \simeq 10^{-10} \, \mathrm{eV}^2 $,
similar in magnitude to those obtained
in the two-generation analyses.

Therefore,
we conclude that the maximally-mixed
three generations of neutrinos are capable
of explaining
both the current atmospheric
and solar neutrino problems
(though marginally in the case of
the Gallium solar neutrino experiments).
Improved data in the future
will decide the fate of this scheme.

Finally,
the obvious question
is:
Is it possible for completely non-degenerate neutrinos
(with the standard mass hierarchy
$ m_{1} \ll  m_{2} \ll m_{3} $)
to have maximal
or, at least, very large mixing?
The conventional wisdom tells us that mixing angles are determined by
lepton mass ratios.
In the quark sector,
the Cabibbo angle is well reproduced by the formula
$\tan^2\theta_{_C}=m_d/m_s$.
In the same spirit, the neutrino mixing angle,
in the case of two generations,
is expected to be determined by the ratios
$m_e/m_\mu$ and $m(\nu_1)/m(\nu_2)$.
If
$ m(\nu_1) \ll m(\nu_2) $,
one expects the neutrino mixing angle
to be very small.

However, this conclusion is not valid when one invokes the see-saw mechanism.
The possibility that the see-saw mechanism may enhance lepton mixing up
to maximal was discussed in
Refs.\cite{B:SMIRNOV93,B:CWKIM93}.
When the see-saw mechanism is invoked with more than one generation of
neutrinos,
the mixing angles
for the resultant light Majorana neutrinos become dependent
upon the masses of heavy right-handed Majorana neutrinos which are usually
provided by GUTS models.
In particular, when certain relationships among the mass parameters
in the original Dirac and heavy right-handed Majorana mass matrices
are satisfied,
the mixing angles for the light Majorana neutrinos can substantially be
enhanced,
even in the case of the usual hierarchy
$ m(\nu_1) \ll m(\nu_2) \ll m(\nu_3) $.

A full analysis of the atmospheric
neutrino problem
with {\it arbitrary} 3 mixing angles
will be given elsewhere.

\acknowledgments

The authors would like to
thank Jaeyoung Lee for his contributions
in the early stage of this work.
They would also like to express their
gratitude to S. Bilenky, A. Bottino and Jewan Kim
for many valuable discussions.
This work was supported in part
by the
Research Funds of the
Ministero dell'Universit\`a
e della Ricerca Scientifica e Tecnologica
and by the
National Science Foundation, USA.

\end{document}